\documentclass[aps,pre,preprint,showpacs,eqsecnum,amssymb]{revtex4}
\usepackage{graphicx,color,psfrag}
\newcommand{\pq}[1]{\left[{#1}\right]}

\begin{document}
\title{Heat flow in chains driven by thermal noise}
\author{Hans C. Fogedby}
\email{fogedby@phys.au.dk}

\affiliation{Department of Physics and Astronomy, University of
Aarhus, Ny Munkegade\\
8000 Aarhus C, Denmark\\}

\affiliation{Niels Bohr Institute, Blegdamsvej 17\\
2100 Copenhagen {\O}, Denmark}

\author{Alberto Imparato}
\email{imparato@phys.au.dk} \affiliation{Department of Physics and
Astronomy\\
University of Aarhus, Ny Munkegade\\
8000 Aarhus C, Denmark}

\begin{abstract}
We consider the large deviation function for a classical harmonic
chain composed of $N$ particles driven at the end points by heat
reservoirs, first derived in the quantum regime by Saito and Dhar
\cite{Saito07} and in the classical regime by Saito and Dhar
\cite{Saito10b} and Kundu et al. \cite{Kundu11}. Within a Langevin
description we perform this calculation on the basis of a standard
path integral calculation in Fourier space. The cumulant
generating function yielding the large deviation function is given
in terms of a transmission Green's function and is consistent with
the fluctuation theorem. We find a simple expression for the tails
of the heat distribution which turn out to decay exponentially.
We, moreover, consider an extension of a single particle model
suggested by Derrida and Brunet \cite{Derrida05} and discuss the
two-particle case.  We also discuss the limit for large $N$ and
present a closed expression for the cumulant generating function.
Finally, we present a derivation of the fluctuation theorem on the
basis of a Fokker-Planck description. This result is not
restricted to the harmonic case but is valid for a general
interaction potential between the particles.
\end{abstract}
.

\maketitle

\section{\label{intro} Introduction}
There is a current interest in the thermodynamics and statistical
mechanics of fluctuating systems in contact with heat reservoirs
and driven by external forces. The current focus stems from the
recent possibility of direct manipulation of nano-systems and
bio-molecules. These techniques permit direct experimental access
to the probability distribution functions for the work or for the
heat exchanged with the environment \cite{Trepagnier04,Collin05,
Seifert06a,Seifert06b,Wang02,Imparato06a,Imparato07,Ciliberto06,
Ciliberto07,Ciliberto08,Imparato08}. These methods have also
yielded access to the experimental verification of the recent
fluctuation theorems which relate the probability of observing
entropy-generated trajectories with that of observing
entropy-consuming trajectories \cite{Jarzynski97,Kurchan98,
Gallavotti96,Crooks99,Crooks00,Seifert05a,Seifert05b,
Evans93,Evans94,Gallavotti95,Lebowitz99,Gaspard04,Imparato06,
vanZon03,vanZon04,vanZon03a,vanZon04a,Seifert05c}.

In recent works  we studied the motion of a Brownian particle in a
general potential with a view to the distribution function for the
heat exchange with the surroundings \cite{Fogedby09a} and a single
bound Brownian particle driven by two heat reservoirs
\cite{Fogedby11a}. In the present paper we consider the harmonic
chain driven by heat reservoirs at temperatures $T_1$ and $T_N$
\cite{Rieder67,Casher71,Nakazawa68,Nakazawa70,Dhar01,Dhar06,
Roy08,Lepri03,Dhar08}. Here the distribution of positions and
momenta is given by a Gaussian form with a correlation matrix with
elements given by the static position and momentum correlations
\cite{Rieder67}.

Owing to the current interest in fluctuation theorems the linear
chain has recently been addressed again by Saito and Dhar
\cite{Saito10b} and by Kundu et al. \cite{Kundu11}; see also
\cite{Saito07} for a treatment in the quantum regime. Using a path
integral formulation, Fourier series, and analyzing the resulting
energy transmission matrices, these authors derive an expression
for the cumulant generating function, in the following denoted
CGF, for the heat transfer in terms of a transmission Green's
function $T(\omega)$. The large deviation function for the heat
transfer, denoted LDF, then follows by a Legendre transformation
of the CGF; for definitions, see later. The expression is in
accordance with the fluctuation theorem
\cite{Evans93,Evans94,Gallavotti95,Lebowitz99,
Kurchan98,Crooks99,Seifert05a}.

In the present paper we consider four issues: i) the CGF for the
harmonic chain, ii) the LDF for the chain and the exponential
tails in the heat distribution, iii) the CGF for an extension of a
model by Derrida and Brunet \cite{Derrida05}, iv) the CGF for the
harmonic chain in the large $N$ limit, where $N$ is the number of
particles, and v) a derivation of the fluctuation theorem on the
basis of a Fokker-Planck description.

For the purpose of the analysis in ii) - iv) we have within a
Langevin scheme performed a calculation of the CGF, including
explicit expressions for the transmission Green function. At the
technical level we, moreover, unlike Kundu et al. \cite{Kundu11},
make use of Fourier transforms throughout the calculation and
diagonalize explicitly $T(\omega)$ expressing the CGF in terms of
the eigenvalues. For the benefit of the reader and the continuity
of the paper we have chosen to include this analysis in the main
part of the paper.

We discuss the tails in the heat distribution and exemplify this
feature both for the extended Derrida-Brunet model and the $N$
particle chain. We consider the CGF and LDF for an extension of a
single particle model suggested by Derrida and Brunet
\cite{Derrida05} and the CGF in the two-particle case. We,
moreover, analyze the asymptotic large $N$ limit and present a
closed expression for the CGF.

Finally, as a related and more formal issue we present a
derivation of the fluctuation theorem on the basis of a
Fokker-Planck description of a chain. As a bonus we are able to
prove that the fluctuation theorem holds for chains with general
interaction potentials and with several heath baths at different
temperatures.

For reference we present below the results of Saito and Dhar
\cite{Saito10b} and Kundu et al. \cite{Kundu11}, also presented in
the present paper. Denoting the model-dependent transmission
Greens function by $T(\omega)$, the CGF $\mu(\lambda)$ for the
characteristic function for the transferred heat $Q(t)$ in the
time interval $t$, is given by the following expressions:
\begin{eqnarray}
&&\langle\exp(\lambda Q(t))\rangle=\exp(t\mu(\lambda)), \label{1}
\\
&&\mu(\lambda)=-\frac{1}{2}\int\frac{d\omega}{2\pi}
\ln[1+T(\omega)f(\lambda)], \label{2}
\\
&&f(\lambda)=T_1T_N\lambda(1/T_1-1/T_N-\lambda). \label{3}
\end{eqnarray}
Here $T_1$ and $T_N$ denote the reservoir temperatures and the
form of $f(\lambda)$ ensures the validity of the
fluctuation theorem
\begin{eqnarray}
\mu(\lambda)=\mu(1/T_1-1/T_N-\lambda). \label{ft1}
\end{eqnarray}

As a new result we present below the CGF in the asymptotic large
$N$ limit. Here $\Gamma$ denotes the reservoir damping and
$\kappa$ the spring constant. The CGF is given by
\begin{eqnarray}
\mu(\lambda)= -\int_0^\pi\frac{dp}{2\pi}\sqrt{\kappa}\cos(p/2)
\ln\left[1+\frac{8\Gamma\kappa^{-1/2}\sin(p/2)\sin(p)f(\lambda)}{1+
4(\Gamma^2/\kappa)\sin^2(p/2)}\right]. \label{4}
\end{eqnarray}
Finally, we note that the mathematical background for the present
paper is provided by large deviation theory, see Refs.
\cite{Touchette09,Freidlin98,Ellis99,Varadhan08,Hollander00}.

The paper is organized in the following manner. In
Sec.~\ref{chain} we present the harmonic chain. In
Sec.~\ref{analysis} we set up the necessary analysis. In
Sec.~\ref{large} we present a derivation of the CGF. The general
properties of such a function are discussed in Sec.~\ref{discuss},
where we also consider the tails of the heat distribution, the
specific cases of a bound Brownian particle, a two-particle chain,
and the large $N$ limit. In Sec.~\ref{gen} we discuss a
generalization of the fluctuation theorem. In Sec.~\ref{sum} we
present a summary and a conclusion.
\section{\label{chain} Harmonic chain}
The dynamics of a unit mass harmonic chain composed of $N$
particles and, moreover, attached to a wall or substrate, is
governed by the Hamiltonian
\begin{eqnarray}
H=\frac{1}{2}\sum_{n=1}^Np_n^2+
\frac{\kappa}{2}\sum_{n=1}^{N-1}(u_n-u_{n+1})^2
+\frac{\kappa}{2}(u_1^2+u_N^2), \label{ham}
\end{eqnarray}
where $u_n$ and $p_n$ denotes displacements and momenta,
respectively; $\kappa$ is the spring constant.  The equation of
motion for the bulk particles and the end particles driven by the
heat reservoirs at temperatures $T_1$ and $T_N$ with associated
damping $\Gamma$ are given by
\begin{eqnarray}
&&\frac{du_n}{dt}=p_n, \label{eq1}
\\
&&\frac{dp_n}{dt}=\kappa(u_{n+1}+u_{n-1}-2u_n), ~~n=2,\cdots N-1,
\label{eq2}
\\
&&\frac{dp_1}{dt}=-\Gamma p_1+\kappa(u_2-2u_1)+\xi_1, \label{eq3}
\\
&&\frac{dp_N}{dt}=-\Gamma p_N+\kappa(u_{N-1}-2u_N)+\xi_N,
\label{eq4}
\end{eqnarray}
with noise correlations and strengths
\begin{eqnarray}
&&\langle\xi_1(t)\xi_1(t')\rangle=\Delta_1\delta(t-t'), \label{n1}
\\
&&\langle\xi_N(t)\xi_N(t')\rangle=\Delta_N\delta(t-t'), \label{n2}
\\
&&\Delta_1=2\Gamma T_1, \label{s1}
\\
&&\Delta_N=2\Gamma T_N;\label{s2}
\end{eqnarray}
in equilibrium $\Delta_1=\Delta_N=\Delta$ and detailed balance
implies $\Delta=2\Gamma T$, where $T$ is the common temperature of
the reservoirs.

Focussing on the reservoir at temperature $T_1$ the fluctuating
force is given by $-\Gamma p_1+\xi_1$ and, correspondingly, the
rate of work or heat flux has the form, denoting $Q\equiv Q_1$,
\begin{eqnarray}
\frac{dQ}{dt}=p_1(-\Gamma p_1+\xi_1). \label{hf}
\end{eqnarray}
The central quantity in the analysis is, however, the total heat
transmitted during a finite time interval $t$, i.e.,
\begin{eqnarray}
Q(t)=\int_0^td\tau p_1(\tau)(-\Gamma p_1(\tau)+\xi_1(\tau));
\label{h}
\end{eqnarray}
note that strictly speaking only the time scaled heat $Q(t)/t$ has
large deviation properties, see e.g. Refs.
\cite{Touchette09,Hollander00}.

\begin{figure}
\includegraphics[width=1.0\hsize]{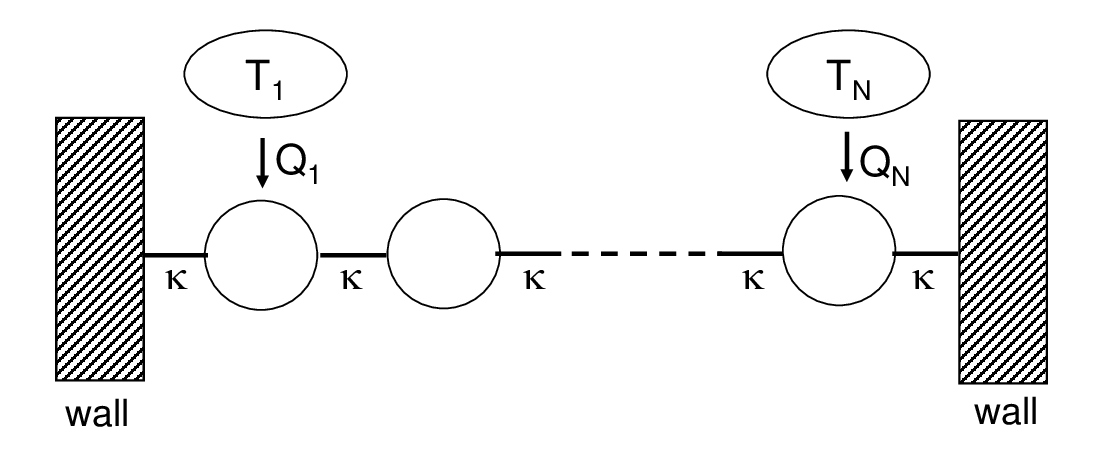}
\caption{We depict a harmonic chain in contact with heat
reservoirs at temperatures $T_1$ and $T_N$. The chain is attached
to walls or substrates at the ends. The total heat transmitted to
the $n=1$ and $n=N$ particles are denoted $Q_1$ and $Q_N$,
respectively. The spring constant is denoted $\kappa$.}
\label{fig1}
\end{figure}

The heat $Q(t)$ is fluctuating and the issue is to determine its
probability distribution $P(Q,t)=\langle\delta(Q-Q(t))\rangle$;
here $\langle\cdots\rangle$ denotes an average with respect to
$\xi_1$ and $\xi_N$. In terms of the characteristic function
$\langle\exp(\lambda Q(t)\rangle$ we have by a Laplace transform
\cite{Gradshteyn65}
\begin{eqnarray}
P(Q,t)=\int_{-i\infty}^{i\infty}\frac{d\lambda}{2\pi i}e^{-\lambda
Q}\langle e^{\lambda Q(t)}\rangle. \label{dh}
\end{eqnarray}
The chain attached to the substrate at the ends and driven by heat
reservoirs is depicted in Fig.~\ref{fig1}.

\section{\label{analysis}Analysis}
The heat reservoirs drive the chain into a stationary state. Since
the heat is transported ballistically the only damping mechanism
is associated with the heat reservoirs and the only time scale is
given by $1/\Gamma$. Consequently, at long times compared with
$1/\Gamma$ we can neglect the initial preparation of the chain and
analyze the problems in terms of Fourier transforms. Thus
introducing the Fourier transform
\begin{eqnarray}
u_n(t)=\int\frac{d\omega}{2\pi}e^{-i\omega t}u_n(\omega),
\label{fot}
\end{eqnarray}
the equations of motion (\ref{eq1}) to (\ref{eq4}) and noise
correlations  (\ref{n1}) to (\ref{n2}) take the form
\begin{eqnarray}
&&\sum_{m=1}^N G^{-1}_{nm}(\omega)u_m(\omega)=
\delta_{n1}\xi_1(\omega)+\delta_{nN}\xi_N(\omega), \label{eq}
\\
&& \langle\xi_1(\omega)\xi_1(\omega')\rangle=
2\pi\Delta_1\delta(\omega+\omega'), \label{n11}
\\
&&\langle\xi_N(\omega)\xi_N(\omega')\rangle=
2\pi\Delta_N\delta(\omega+\omega'). \label{n22}
\end{eqnarray}
Here the inverse Green's function $G^{-1}_{nm}(\omega)$ is a
symmetrical tridiagonal matrix with elements
\begin{eqnarray}
&&G_{11}^{-1}(\omega)= G_{NN}^{-1}(\omega)=\Omega,\label{g1}
\\
&&G_{nn}^{-1}(\omega)=\widetilde{\Omega},~~n=2,\cdots
N-1,\label{g2}
\\
&&G_{nn+1}^{-1}(\omega)=G_{nn-1}^{-1}(\omega)= -\kappa, \label{g3}
\end{eqnarray}
where
\begin{eqnarray}
&&\Omega = -\omega^2+2\kappa-i\Gamma\omega, \label{ff1}
\\
&&\widetilde{\Omega} = -\omega^2+2\kappa; \label{ff2}
\end{eqnarray}
note that for a free chain we have
$\Omega=-\omega^2+\kappa-i\Gamma\omega$.

Propagating bulk solutions have the form
\begin{eqnarray}
&&u_n(\omega)=A\exp(ipn)+B\exp(-ipn), \label{u}
\\
&&\omega^2=4\kappa\sin^2(p/2),~~ \label{dis}
\end{eqnarray}
where $p$ is confined to the first Brillouin zone $|p|<\pi$ and,
correspondingly, $|\omega|<2\sqrt{\kappa}$. Imposing the noisy
drive we readily determine the coefficients $A$ and $B$ and infer
the solutions
\begin{eqnarray}
u_n(\omega)=G_{n1}(\omega)\xi_1(\omega)+G_{nN}(\omega)\xi_N(\omega),
~~p_n(\omega)=(-i\omega)u_n(\omega), \label{sol}
\end{eqnarray}
where the Green function components are given by
\begin{eqnarray}
&&G_{n1}(\omega)=\frac{\Omega\sin(N-n)p-\kappa\sin(N-n-1)p}
{D(\omega)},\label{G1}
\\
&&G_{nN}(\omega)=\frac{\Omega\sin(n-1)p-\kappa\sin(n-2)p}{D(\omega)},
\label{GN}
\\
&&D(\omega)=
\Omega^2\sin(N-1)p-2\kappa\Omega\sin(N-2)p+\kappa^2\sin(N-3)p.
\label{d}
\end{eqnarray}
The displacement $u_n$ is thus driven by stochastically excited
lattice waves (phonons) propagating towards the site from the end
points; $D(\omega)=0$  yield the damped mode spectrum. Also, from
the definition of $G_{nm}^{-1}(\omega)$ we deduce the relationship
$G_{nm}^{-1}(\omega)-G_{nm}^{-1}(-\omega)=
-2i\omega\Gamma\delta_{nm}(\delta_{n1}+\delta_{nN})$ and by
multiplication the Schwinger identity \cite{Wang45}
\begin{eqnarray}
G_{nm}(\omega)-G_{nm}(\omega)^\ast=
2i\omega\Gamma[G_{n1}(\omega)G_{1m}(\omega)^\ast+
G_{nN}(\omega)G_{Nm}(\omega)^\ast].\label{sc}
\end{eqnarray}

In the absence of the heat reservoirs  energy is conserved, i.e.,
$dH/dt=0$, where $H$ is given by (\ref{ham}). Coupling the
reservoirs to the chain we have $dH/dt=dQ_1/dt+dQ_N/dt$, where
$dQ_N/dt$ is the heat flux from the reservoir at temperature
$T_N$. Averaging we have for the mean heat fluxes $\langle
dQ_1/dt\rangle= -\langle dQ_N/dt\rangle$, expressing the energy
balance; the mean input flux at $n=1$ is equal to the mean output
flux at $n=N$.

Using (\ref{hf}), inserting (\ref{sol}), averaging over the noises
(\ref{n11}) and (\ref{n22}), using the properties of the Green's
function (\ref{G1}) and (\ref{GN}), and the identity (\ref{sc}),
we obtain for the mean transferred heat in time $t$
\begin{eqnarray}
\langle Q(t)\rangle=t(\Delta_1-\Delta_N)\Gamma
\int\frac{d\omega}{2\pi}\omega^2|G_{1N}(\omega)|^2. \label{mean}
\end{eqnarray}
Here the central model dependent quantity is the end-to-end Greens
function $G_{1N}(\omega)$; we note that the mean heat vanishes for
$\Delta_1=\Delta_N$. We also note that the transferred mean heat
rate $\bar q=\langle Q\rangle/t$ is given by
\begin{equation}
\bar q=\Gamma (T_1-\langle p_1^2\rangle); \label{qbar}
\end{equation}
see Ref. \cite{Rieder67}. Here $\langle p_1^2\rangle$ is the
average kinetic temperature of the first particle in the steady
state; note that in equilibrium $\bar q=0$ and $\langle
p_1^2\rangle=T_1$ in accordance with the equipartition theorem
\cite{Reichl98}. The relation (\ref{qbar}) follows from
(\ref{mean}) by inserting the Greens function solution of the
equations of motion and using the identity (\ref{sc}).

The expression (\ref{qbar}) also follows from the equivalent
Fokker-Planck approach to the harmonic chain which we discuss
below. Considering the definition of the n-th moment of the heat
transfer in time $t$
\begin{eqnarray}
\langle Q^n(t)\rangle=\int dQ du dp~ Q^nP(u,p,Q,t), \label{mom}
\end{eqnarray}
and referring to (\ref{fp2}) in Sec.~\ref{gen}, the Fokker-Planck
equation for the joint distribution $P(u,p,Q,t)$ in the case of
two reservoirs implies
\begin{eqnarray}
\frac{d\langle Q^n\rangle}{dt}= \Gamma
n\langle(T_1-p_1^2)Q^{n-1}\rangle+\Gamma n(n-1) T_1\langle p_1^2
Q^{n-2}\rangle. \label{momeq}
\end{eqnarray}
These equations of motion are part of a hierarchy relating the
n-th moment to correlations of the lower moments with $\langle
p_1^2\rangle$ and have to be completed by equations of motions for
the correlations $\langle p_1^2 Q^{n-2}\rangle$. Without further
assumptions this hierarchy will in general not terminate and
simply represents a reformulation. We note, however, that for the
first moment for $n=1$ the second term in (\ref{momeq}) vanishes
and we obtain a closed equation yielding (\ref{qbar}).

For the fluctuating heat transferred in time $t$ we obtain, using
(\ref{h}) and inserting (\ref{sol}), the expression
\begin{eqnarray}
Q(t)=\int\frac{d\omega}{2\pi}\frac{d\omega'}{2\pi}
F(\omega-\omega') \left(
  \begin{array}{cc}
    \xi_1(\omega) & \xi_N(\omega) \\
  \end{array}
\right)M(\omega,\omega') \left(
  \begin{array}{c}
    \xi_1(-\omega')\\
    \xi_N(-\omega')\\
  \end{array}
\right). \label{heat}
\end{eqnarray}
The heat transfer is a fluctuating quantity depending bilinearly
on the reservoir noises $\xi_1$ and $\xi_N$. The matrix elements
in the symmetrical form (\ref{heat}) are given by
\begin{eqnarray}
&&M_{11}(\omega,\omega')=-\Gamma
A(\omega)A(\omega')^\ast+(1/2)(A(\omega)+A(\omega')^\ast),
\label{m1}
\\
&&M_{22}(\omega,\omega')=-\Gamma B(\omega)B(\omega')^\ast,
\label{m2}
\\
&&M_{12}(\omega,\omega')=-\Gamma
A(\omega)B(\omega')^\ast+(1/2)B(\omega')^\ast, \label{m3}
\\
&&M_{21}(\omega,\omega')=-\Gamma
B(\omega)A(\omega')^\ast+(1/2)B(\omega), \label{m4}
\end{eqnarray}
where we have introduced the
notation
\begin{eqnarray}
&&A(\omega)=-i\omega G_{11}(\omega), \label{a1}
\\
&&B(\omega)=-i\omega G_{1N}(\omega); \label{a2}
\end{eqnarray}
we note that (\ref{sc}) implies
\begin{eqnarray}
A(\omega)+A(\omega)^\ast=2\Gamma[|A(\omega)|^2+|B(\omega)|^2].
\label{sc2}
\end{eqnarray}
The dependence on the transfer time $t$ is embodied in the
function
\begin{eqnarray}
F(\omega)= 2e^{-i\omega t/2}\frac{\sin(\omega t/2)}{\omega}.
\label{f}
\end{eqnarray}
For later purposes we also note that
\begin{eqnarray}
&&F(0)=t, \label{f1}
\\
&&|F(\omega)|^2=2\pi t\delta(\omega)~~\text{for large}~t.
\label{f2}
\end{eqnarray}
At this stage our calculation differs from Kundu et al.
\cite{Kundu11} in that we use a Fourier transform instead of a
Fourier series in the expression (\ref{heat}) for the fluctuating
heat. The dependence on the transfer time $t$ is then incorporated
in the function $F(\omega)$.
\section{\label{large}Large deviation function}
For large $t$ the mean heat $\langle Q(t)\rangle$ given by
(\ref{mean}) grows linearly with time. Analyzing the higher
cumulants $\langle Q(t)^n\rangle_c$, i.e., $\langle
Q(t)^2\rangle_c=\langle Q(t)^2\rangle-\langle Q(t)\rangle^2$,
etc., by averaging over the noise and applying Wick's theorem
\cite{Zinn-Justin89}, it also follows that they likewise increase
linearly with time, i.e., $\langle Q(t)^n\rangle_c\sim t$ for
large $t$. We thus infer from the cumulant expansion of the
characteristic function \cite{Reichl98},
\begin{eqnarray}
\langle \exp(\lambda
Q(t))\rangle=\exp\left(\sum_{n=0}^\infty\frac{\lambda^n}{n!}
\langle Q(t)^n\rangle_c\right), \label{cum}
\end{eqnarray}
that for large $t$
\begin{eqnarray}
\langle \exp(\lambda Q(t))\rangle= \exp(t\mu(\lambda)),
\label{char}
\end{eqnarray}
where $\mu(\lambda)$ is the cumulant generating function, denoted
CGF.

The CGF characterizes the long time heat distribution. From the
cumulant expansion (\ref{cum}) we obtain the relationship
\begin{eqnarray}
\left(\frac{d^n\mu(\lambda)}{d\lambda^n}\right)_{\lambda=0}=
\frac{\langle Q(t)^n\rangle_c}{t}. \label{cum2}
\end{eqnarray}
Here the definition (\ref{char}) for $\lambda=0$ implies
\begin{eqnarray}
\mu(0)=0. \label{norm}
\end{eqnarray}
In case the fluctuation theorem is valid we, moreover, have the
symmetry
\begin{eqnarray}
\mu(\lambda)=\mu(1/T_1-1/T_N-\lambda). \label{ft}
\end{eqnarray}
Turning to the evaluation of $\mu(\lambda)$ we average
$\exp(\lambda Q(t))$ with respect to the noises $\xi_1$ and
$\xi_N$. In matrix form the Gaussian noise distribution has the
form
\begin{eqnarray}
P(\xi)\propto\exp\left(-\frac{1}{2}
\int\frac{d\omega}{2\pi}\frac{d\omega'}{2\pi}
\tilde\xi(\omega)\Delta^{-1}(\omega-\omega')\xi(-\omega')\right),
\label{ndis}
\end{eqnarray}
where $\tilde\xi(\omega)=(\xi_1(\omega),\xi_N(\omega))$ and the
inverse noise matrix is given by
\begin{eqnarray}
\Delta^{-1}(\omega-\omega')= \left(
  \begin{array}{cc}
    \Delta_1^{-1} & 0 \\
    0 & \Delta_N^{-1} \\
  \end{array}
\right)\delta(\omega-\omega'). \label{nm}
\end{eqnarray}
Noting from (\ref{heat}) that $Q(t)$ is bilinear in $\xi$ and
using the identities \cite{Zinn-Justin89}
\begin{eqnarray}
&&\langle\exp(-(1/2)\tilde\xi B\xi)\rangle= \det(I+\Delta
B)^{-1/2}, \label{id1}
\\
&&\det(A)=\exp(\text{Tr}\ln(A)), \label{id2}
\end{eqnarray}
we obtain for the CGF
\begin{eqnarray}
\mu(\lambda)=-\frac{1}{2t}\text{Tr}\ln(I-2\lambda\Delta FM).
\label{ldf}
\end{eqnarray}
In the remaining part of this section the present calculation
differs from  Kundu et al. \cite{Kundu11} in that we owing to the
nondiagonal character of $F$ must expand $\mu$ in order to
implement the large $t$ limit. Thus expanding the log according to
$\ln(1+x)= \sum_{n=1}(-1)^{n+1}x^n/n$ and tracing term by term we
have
\begin{eqnarray}
\mu(\lambda)=
-\frac{1}{2t}\sum_{n=1}\frac{(-1)^{n+1}}{n}(-2\lambda)^n
\text{Tr}(\Delta FM)^n, \label{ldf2}
\end{eqnarray}
and the issue is to determine $\text{Tr}(\Delta FM)^n$ and
complete the sum. From (\ref{heat}) we obtain
\begin{eqnarray}
\text{Tr}(\Delta FM)^n=\int\prod_{k=1}^n\frac{d\omega_k}{2\pi}
F(\omega_k-\omega_{k+1})\text{Tr}\left(\prod_{k=1}^n\Delta
M(\omega_k,\omega_{k+1})\right), \label{tr}
\end{eqnarray}
where $\omega_{n+1}=\omega_1$. Inserting (\ref{f}) we notice that
since $\sum_{k=1}^n(\omega_k-\omega_{k+1})=0$ the exponential
factors in the product of F functions combine yielding a unit
factor. We thus only have to retain the sine part, i.e.,
$F(\omega)\rightarrow 2\sin(\omega t/2)/\omega$. Using (\ref{f1})
and (\ref{f2}) we have for $n=1,2$
\begin{eqnarray}
&&\text{Tr}(\Delta FM)=t\int\frac{d\omega}{2\pi}\text{Tr}(\Delta
M(\omega,\omega)), \label{tr2}
\\
&&\text{Tr}(\Delta FM)^2=t\int\frac{d\omega}{2\pi}
\text{Tr}(\Delta M(\omega,\omega)\Delta M(\omega,\omega)).
\label{tr3}
\end{eqnarray}
For large $t$ the function $F(\omega)$ oscillates rapidly as a
function of $\omega$ and we have approximately
$\omega_1\sim\omega_2\cdots\sim\omega_n$, i.e., the effective
integration range in $\omega$ space is confined to the domain
$\omega_1=\omega_2=\cdots=\omega_n$ and only one $\omega$
integration remains. Using $\int(d\omega/2\pi)F(\omega)=1$
inspection readily yields
\begin{eqnarray}
\text{Tr}(\Delta FM)^n= t\int\frac{d\omega}{2\pi}\text{Tr}((\Delta
M(\omega,\omega)^n), \label{trn}
\end{eqnarray}
and the CGF takes the form
\begin{eqnarray}
\mu(\lambda)=-\frac{1}{2}
\sum_{n=1}\frac{(-1)^{n+1}}{n}(-2\lambda)^n
\int\frac{d\omega}{2\pi}\text{Tr}((\Delta M(\omega,\omega)^n).
\label{ldf3}
\end{eqnarray}
In order to complete the calculation we diagonalize the two-by-two
matrix $\Delta M$. Denoting the eigenvalues by $\alpha_+(\omega)$
and $\alpha_+(\omega)$ we have $\text{Tr}((\Delta
M(\omega,\omega)^n)= \alpha_+(\omega)^n+\alpha_-(\omega)^n$ and
reconstructing the log we obtain for $\mu(\lambda)$
\begin{eqnarray}
\mu(\lambda)=-\frac{1}{2}\int\frac{d\omega}{2\pi}
[\ln(1-2\lambda\alpha_+(\omega))+\ln(1-2\lambda\alpha_-(\omega))],
\label{ldf33}
\end{eqnarray}
The eigenvalues $\alpha_+(\omega)$ and $\alpha_+(\omega)$ are
determined by the condition $\det(\Delta M-\alpha I)=0$, i.e.,
\begin{eqnarray}
\left|
\begin{array}{cc}
\Delta_1M_{11}(\omega,\omega)-\alpha(\omega)&\Delta_1M_{12}
(\omega,\omega)
\\
\Delta_NM_{21}(\omega,\omega)&\Delta_NM_{22}
(\omega,\omega)-\alpha(\omega)
\\
\end{array}
\right|=0, \label{det}
\end{eqnarray}
yielding the quadratic equation
\begin{eqnarray}
\alpha^2-
\alpha(\Delta_1M_{11}+\Delta_NM_{22})+\Delta_1\Delta_N(M_{11}M_{22}
-M_{12}M_{21})=0, \label{qeq}
\end{eqnarray}
with roots $\alpha_+$ and $\alpha_-$. In particular
\begin{eqnarray}
&&\alpha_++\alpha_-=\Delta_1M_{11}+\Delta_NM_{22}\label{eig1},
\\
&&\alpha_+\alpha_-=\Delta_1\Delta_N(M_{11}M_{22}-M_{12}M_{21})
\label{eig2}.
\end{eqnarray}
Using the identity (\ref{sc2}) we obtain the reduced expressions
\begin{eqnarray}
&&M_{11}(\omega,\omega)=-M_{22}(\omega,\omega)=\Gamma|B(\omega)|^2,
\label{r1}
\\
&&M_{12}(\omega,\omega)=M_{21}(\omega,\omega)^\ast=-\Gamma
A(\omega)B(\omega)^\ast+(1/2)B(\omega)^\ast \label{r2},
\end{eqnarray}
i.e.,
\begin{eqnarray}
&&\alpha_++\alpha_-= (\Delta_1-\Delta_N)\Gamma|B|^2, \label{eig11}
\\
&&\alpha_+\alpha_-=-\Delta_1\Delta_N|B|^2/4, \label{eig22}
\end{eqnarray}
and for the CGF
\begin{eqnarray}
\mu(\lambda)=-\frac{1}{2}\int\frac{d\omega}{2\pi}
\ln\left[1-2\lambda(\Delta_1-\Delta_N)\Gamma|B(\omega)|^2-
\lambda^2\Delta_1\Delta_N|B(\omega)|^2\right]. \label{ldf6}
\end{eqnarray}
Finally, inserting (\ref{s1}) and (\ref{s2}) the CGF can be
expressed in the form
\begin{eqnarray}
\mu(\lambda)= -\frac{1}{2}\int\frac{d\omega}{2\pi}
\ln[1+4\Gamma^2|B(\omega)|^2f(\lambda)], \label{ldff}
\end{eqnarray}
where
\begin{eqnarray}
&&B(\omega)=-i\omega G_{1N}(\omega) \label{a22},
\\
&&f(\lambda)=T_1T_N\lambda(-\lambda+1/T_1-1/T_N). \label{ff}
\end{eqnarray}
This expression is in agreement with Kundu et al. \cite{Kundu11}.
Here the form of $f(\lambda)$ ensures that the fluctuation theorem
(\ref{ft}) holds. The deterministic dynamics of the chain is
entirely embodied in the momentum Green's function $B(\omega)$.
\section{\label{discuss}Discussion}
Here we discuss four issues: i) the branch cut structure in
$\mu(\lambda)$ and ensuing exponential tails in the heat
distribution $P(Q/t)$, ii) a single bound Brownian particle
coupled to two reservoirs, iii) a two particle chain coupled to
heat reservoirs, and iv) an asymptotic expression for
$\mu(\lambda)$ in the large $N$ limit.
\subsection{Exponential tails}
\begin{figure}[h]
\center \psfrag{mu}[ct][ct][1.5]{$\mu(\lambda)$}
\psfrag{l}[ct][ct][1.5]{$\lambda$}
\psfrag{lmin}[ct][ct][1.5]{$\lambda_-$}
\psfrag{lmax}[ct][ct][1.5]{$\lambda_+$}
\includegraphics[width=15cm]{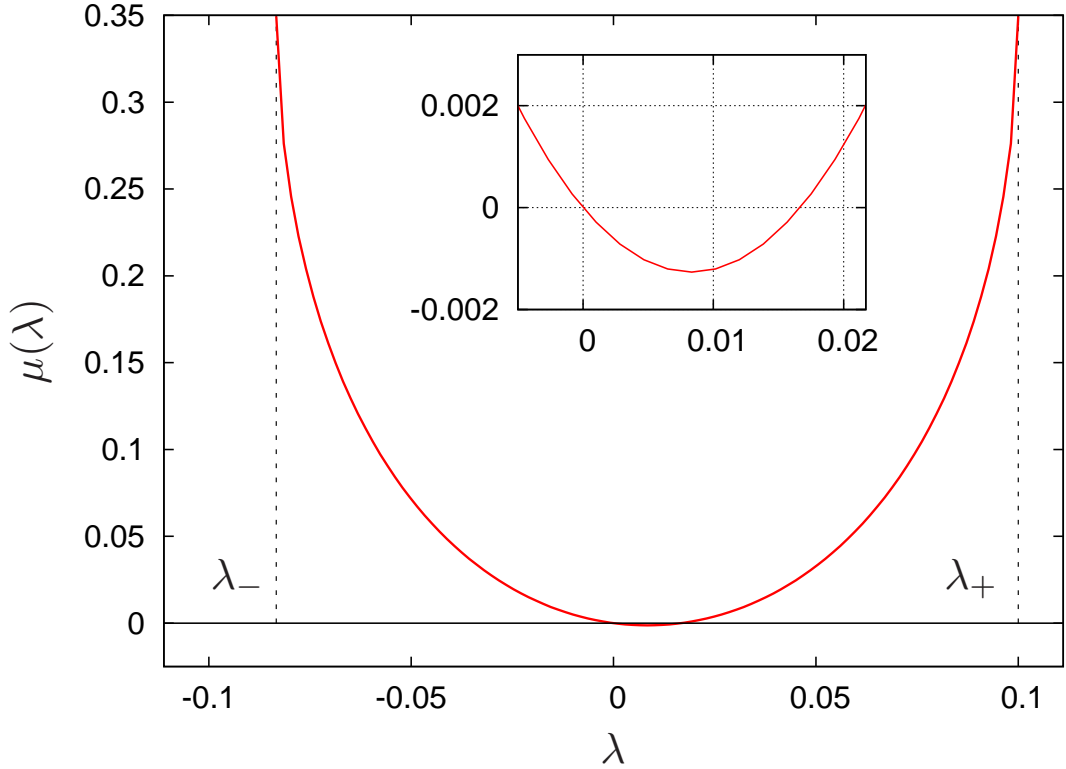}
\caption{Cumulant generating function $\mu(\lambda)$, as given by
(\ref{ldff}) for $T_1=10$, $T_N=12$, $\Gamma=2$, $\kappa=1$,
$N=10$. Inset: zoom of the plot for small value of $\lambda$.}
\label{fig2}
\end{figure}

By inspection of the general expression (\ref{ldff}) for the CGF
we infer that $\mu(\lambda)$ has the form of a downward convex
function passing through the origin $\mu(0)=0$ due to
normalization and through $\mu(1/T_1-1/T_N)=0$ owing to the
fluctuation theorem. Since the argument in the log in (\ref{ldff})
must be positive we infer the condition
\begin{eqnarray}
f(\lambda)\geq -\frac{1}{4\Gamma^2|B|^2_{\text{max}}},
\label{cond}
\end{eqnarray}
where $|B|_{\text{max}}$ is the maximum value of $|B(\omega)|$ in
the $\omega$ range. By means of algebraic and trigonometric
manipulations it can be shown that $|B(\omega)|^2$ is bounded by $
1/4\Gamma^2$, for details see appendix \ref{app1}, and
consequently, $f(\lambda)\geq -1$. By analyzing the expression for
$f(\lambda)$ in (\ref{ff}) one easily finds that this bound is
satisfied for $\lambda_-\le\lambda\le \lambda_+$, where the branch
points $\lambda_\pm$ in $\mu(\lambda)$ are  given by
\begin{eqnarray}
&&\lambda_+=1/T_1, \label{brp1}
\\
&&\lambda_-=1/T_N. \label{brp2}
\end{eqnarray}
In Fig.~\ref{fig2} we have depicted the CGF given by (\ref{ldff})
for the case $T_1=10$, $T_N=12$, $\Gamma=2$, $\kappa=1$, and
$N=10$.

\begin{figure}
\center\psfrag{fq}[ct][ct][1.5]{$-F(q)$}
\psfrag{q}[ct][ct][1.5]{$q$}
\includegraphics[width=15cm]{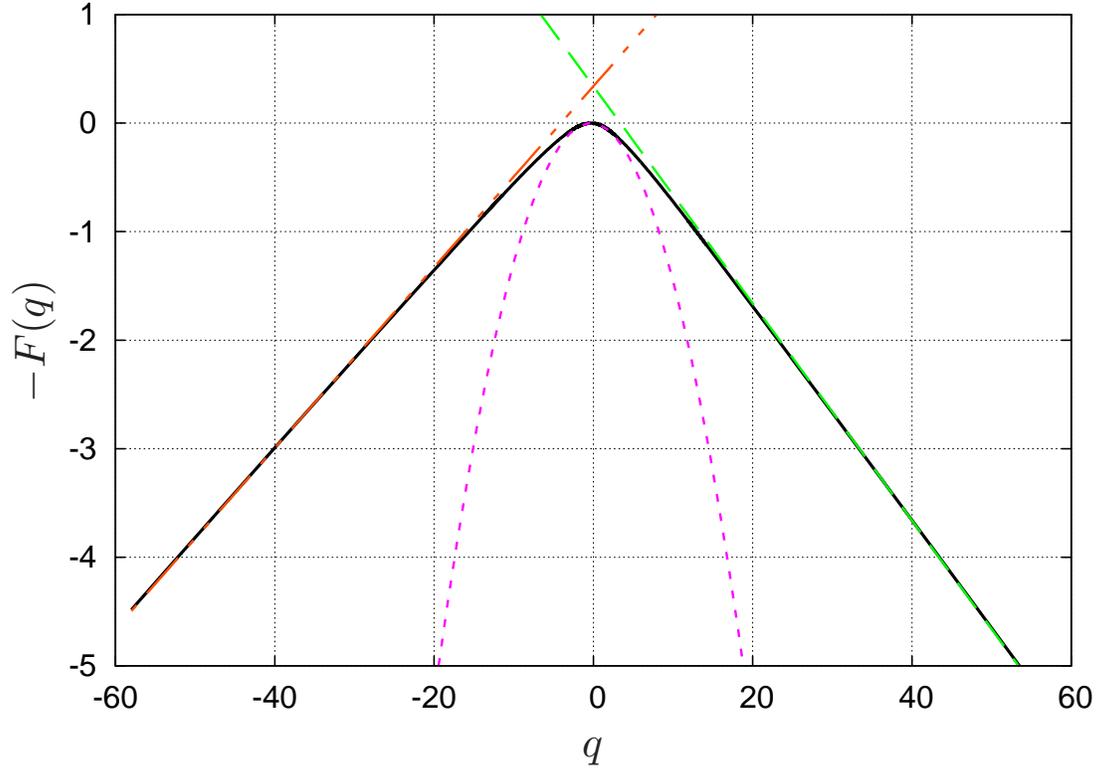}
\caption{Full line: plot of the large deviation function $-F$ as a
function of $q$, as given by (\ref{pldf}) for $T_1=10$, $T_N=12$,
$\Gamma=2$, $\kappa=1$, $N=10$. Dotted line: parabolic
approximation, (\ref{quad2}). Dashed and dotted-dashed line:
Linear regime for $|q|\gg\bar q$, the slopes are $-1/T_1$ and
$1/T_N$, respectively.} \label{fig3}
\end{figure}

At large times the heat distribution function follows from
(\ref{dh}), i.e.,
\begin{eqnarray}
P(Q,t)=\int_{-i\infty}^{i\infty}\frac{d\lambda}{2\pi i}e^{-\lambda
Q}e^{t\mu(\lambda)}, \label{dh2}
\end{eqnarray}
and the rate function or large deviation function $F(q)$ is given
by
\begin{eqnarray}
&&P(q)\sim e^{-tF(q)}, \label{dh4}
\\
&&q=\frac{Q}{t}. \label{scQ}
\end{eqnarray}
Since $\mu(\lambda)$ is differentiable, strictly convex, and steep
at the boundaries the G\"{a}rtner-Ellis theorem
\cite{Touchette09,Hollander00} implies that the LDF is given by
the Legendre transform
\begin{eqnarray}
F(q)= {\sup}_\lambda\{q\lambda-\mu(\lambda)\}, \label{leg}
\end{eqnarray}
or
\begin{eqnarray}
P(q,t) \sim e^{t(\mu(\lambda^\ast)-\lambda^\ast q)}, \label{dh3}
\end{eqnarray}
where $\lambda^\ast$ is determined by
\begin{eqnarray}
\mu'(\lambda^\ast)=q, \label{spc}
\end{eqnarray}
and we find the LDF
\begin{eqnarray}
F(q)=-\mu(\lambda^\ast)+\lambda^\ast\mu'(\lambda^\ast).
\label{pldf}
\end{eqnarray}
For $F(q)$ the fluctuation theorem has the form
\begin{eqnarray}
F(q)-F(-q)=q(1/T_1-1/T_N). \label{ft2}
\end{eqnarray}
Note that the LDF also follows from a heuristic saddle point
argument, see \cite{Lebowitz99}. In Fig.~\ref{fig3} we have
depicted $-F(q)$ for the case $T_1=10$, $T_N=12$, $\Gamma=2$,
$\kappa=1$, and $N=10$.

Replacing $\mu(\lambda)$ by the parabolic approximation
\begin{eqnarray}
\mu_{\mathrm{par}}(\lambda)=\bar q\lambda(T_1T_N\lambda+T_1-T_N),
\label{quad}
\end{eqnarray}
where $\bar q$ is given by (\ref{qbar}) we obtain for $F(q)$
\begin{eqnarray}
F_{\mathrm{par}}(q)=-\frac{(q-\bar q)^2 (T_1-T_N)}{4 \bar q
T_1T_N}, \label{quad2}
\end{eqnarray}
in accordance with (\ref{ft2}). For the heat distribution we
obtain the displaced Gaussian $P(q)\propto\exp(-t
F_{\mathrm{par}}(q))$; this also follows from general large
deviation theory \cite{Touchette09,Hollander00}.

Deforming the contour in the integral (\ref{dh2}) to pass along
the real axis we pick up branch cut contributions in
$\mu(\lambda)$. Heuristically, we conclude that for large $|q|$
\begin{eqnarray}
&&F(q)\sim \lambda_+q,~~~~\text{for}~q\gg 0~, \label{Fp}
\\
&&F(q)\sim |\lambda_-||q|,~\text{for}~q\ll 0 \label{Fn},
\end{eqnarray}
where $\lambda_+$ and $\lambda_-$ have been  defined above. This
also follows directly form the Legendre transformation since
$\mu(\lambda)$ is defined on a compact support. The linear
behavior is confirmed by the plot of $F(q)$ for our particular
choice of the parameter set, see Fig.~\ref{fig3}. The heat
distribution thus exhibits exponential tails for large $|q|$,
i.e.,
\begin{eqnarray}
&&P(q)\propto\exp(-\lambda_+qt)~\text{for}~q\gg 0, \label{pp}
\\
&&P(q)\propto\exp(-|\lambda_-||q|t)~\text{for}~q\ll 0, \label{pn}
\end{eqnarray}
with $\lambda_+$ and $\lambda_-$ given by (\ref{brp1}) and
(\ref{brp2}). It is interesting that the tails are determined only
by the reservoir temperatures. Finally, we note that the
exponential tails in $P(q)$ also follows from large deviation
theory since $\mu$ is bounded by $\lambda_\pm$, see Refs.
\cite{Touchette09,Hollander00}.
\subsection{Bound Brownian particle}
In an interesting paper Derrida and Brunet \cite{Derrida05}
considered a single Brownian particle driven by two reservoirs at
distinct temperatures and presented an explicit expression for the
CGF $\mu(\lambda)$. This toy model has also been discussed by
Visco \cite{Visco06} who considered next leading term and the role
of initial conditions; see also Farago \cite{Farago02}.

\begin{figure}
\includegraphics[width=1.0\hsize]{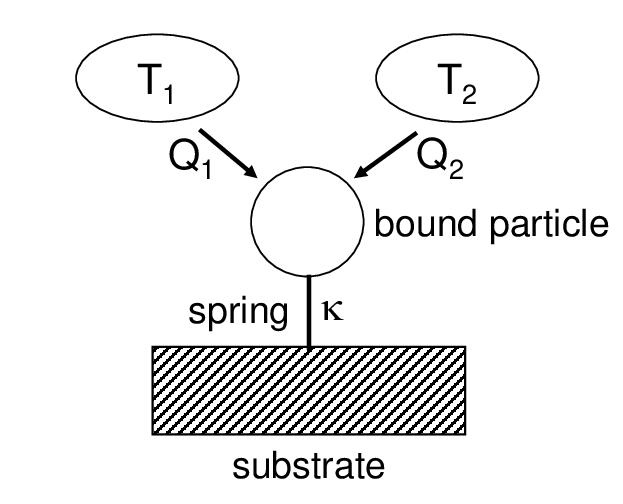}
\caption{We depict a harmonically bound particle interacting with
heat reservoirs at temperatures $T_1$ and $T_2$. The heat
transferred to the particle is denoted $Q_1$ and $Q_2$,
respectively. The particle is attached to a substrate with a
harmonic spring with force constant $\kappa$. } \label{fig4}
\end{figure}

In a previous paper \cite{Fogedby11a} we considered an extension
of this model to the case of a single particle attached
harmonically to a substrate with spring constant $\kappa$ using
the simple method devised by Derrida and Brunet. We found that the
CGF is {\it independent} of $\kappa$, indicating that the
deterministic character of the spring does not influence the
statistical properties of the long time heat transfer. Here we
consider as an illustration the same problem within the present
scheme and recover a CGF independent of $\kappa$. The
configuration is shown in Fig.~\ref{fig4}.

Associating the damping constants $\Gamma_1$ and $\Gamma_2$ with
the two reservoirs the equation of motion take the form
\begin{eqnarray}
&&\frac{du}{dt}=p, \label{eq1b}
\\
&&\frac{dp}{dt}=-(\Gamma_1+\Gamma_2)p-\kappa u +\xi_1+\xi_2,
\label{eq2b}
\end{eqnarray}
with noise correlations
\begin{eqnarray}
&&\langle\xi_1(t)\xi_1(t')\rangle=2\Gamma_1T_1\delta(t-t'),
\label{nb1}
\\
&&\langle\xi_2(t)\xi_2(t')\rangle=2\Gamma_2T_2\delta(t-t').
\label{nb2}
\end{eqnarray}
In Fourier space we obtain the solution
\begin{eqnarray}
p(\omega)=B(\omega)(\xi_1(\omega)+\xi_1(\omega)), \label{solb}
\end{eqnarray}
where
\begin{eqnarray}
B(\omega)=
\frac{-i\omega}{-\omega^2+\kappa-i(\Gamma_1+\Gamma_2)\omega},
\label{ab}
\end{eqnarray}
and
\begin{eqnarray}
|B(\omega)|^2=
\frac{\omega^2}{(\omega^2-\kappa)^2+(\Gamma_1+\Gamma_2)^2\omega^2};
\label{ab2}
\end{eqnarray}
note that $B(\omega)$ satisfies the Schwinger identity
(\ref{sc2}), i.e.,
\begin{eqnarray}
B(\omega)+B(\omega)^\ast=2(\Gamma_1+\Gamma_2)|B(\omega)|^2.
\label{sc3}
\end{eqnarray}
The heat flux from the reservoir at temperature $T_1$ is
\begin{eqnarray}
\frac{dQ}{dt}=p(-\Gamma_1p+p\xi_1), \label{hb}
\end{eqnarray}
and we obtain from (\ref{solb}), (\ref{heat}), and (\ref{sc3}) the
diagonal matrix elements
\begin{eqnarray}
&&M_{11}(\omega,\omega)=~~\Gamma_2|B(\omega)|^2, \label{mb1}
\\
&&M_{22}(\omega,\omega)=-\Gamma_1|B(\omega)|^2, \label{mb2}
\\
&&M_{12}(\omega,\omega)=-\Gamma_1|B(\omega)|^2+
(1/2)B(\omega)^\ast, \label{mb3}
\\
&&M_{21}(\omega,\omega)=-\Gamma_1|B(\omega)|^2+(1/2)B(\omega).
\label{mb4}
\end{eqnarray}
Following the prescription in Sec.~\ref{large} the eigenvalue
equation imply
\begin{eqnarray}
&&\alpha_++\alpha_-= 2\Gamma_1\Gamma_2(T_1-T_2)|B|^2,
\label{eigb1}
\\
&&\alpha_+\alpha_-= -\Gamma_1\Gamma_2T_1T_2|B|^2, \label{eigb2}
\end{eqnarray}
and we obtain the CGF
\begin{eqnarray}
\mu(\lambda)= -\frac{1}{2}\int\frac{d\omega}{2\pi}
\ln\left[1+4\Gamma_1\Gamma_2|B(\omega)|^2f(\lambda)\right],
\label{ldfb}
\end{eqnarray}
where $|B(\omega)|^2$ is given by (\ref{ab2}) and $f(\lambda)$ by
(\ref{ff}).

A straightforward evaluation of (\ref{ldfb}) using the integral
\cite{Gradshteyn65}
\begin{eqnarray}
\int\frac{d\omega}{2\pi}\ln\left(\frac{\omega^2+a^2}
{\omega^2+b^2}\right) =a-b, \label{int}
\end{eqnarray}
yields
\begin{eqnarray}
\mu(\lambda)=(1/2)(a_++a_--b_+-b_-), \label{ldfb2}
\end{eqnarray}
where
\begin{eqnarray}
&&a_\pm^2=(1/2)((\Gamma_1+\Gamma_2)^2-2\kappa\pm
\sqrt{((\Gamma_1+\Gamma_2)^2-2\kappa)^2-4\kappa^2},
\\
&&b_\pm^2=(1/2)(4\Gamma_1\Gamma_2f(\lambda)+
(\Gamma_1+\Gamma_2)^2-2\kappa\pm
\sqrt{(4\Gamma_1\Gamma_2f(\lambda)+(\Gamma_1+
\Gamma_2)^2-2\kappa)^2-4\kappa^2}.~~~~~~~~
\end{eqnarray}
Further inspection shows, however, that the combination
$a_++a_--b_+-b_-$ is {\it independent} of the spring constant
$\kappa$, as already shown in \cite{Fogedby11a}, and we obtain
\begin{eqnarray}
\mu(\lambda)=(1/2)\left[\Gamma_1+\Gamma_2-
\sqrt{(\Gamma_1+\Gamma_2)^2+4\Gamma_1\Gamma_2f(\lambda)}\right].
\label{ldfb3}
\end{eqnarray}
Introducing $f(\lambda)$, as defined in (\ref{ff}), we can express
(\ref{ldfb3}) in the form
\begin{eqnarray}
\mu(\lambda)=\frac{\Gamma_1+\Gamma_2}{2}-\sqrt{\Gamma_1\Gamma_2T_1T_2}
\sqrt{(\lambda_+-\lambda)(\lambda-\lambda_-)}, \label{ldfb4}
\end{eqnarray}
where the branch points are given by
\begin{eqnarray}
\lambda_\pm=\frac{1}{2}\left[1/T_1-1/T_2\pm\sqrt{(1/T_1-1/T_2)^2+
(\Gamma_1+\Gamma_2)^2/\Gamma_1\Gamma_2T_1T_2}\right]. \label{lamb}
\end{eqnarray}
We note that $|B(\omega)|^2$ given by (\ref{ab2}) has a two-peak
structure with maximum value $1/(\Gamma_1+\Gamma_2)^2$ at
$\omega=\pm\sqrt\kappa$ and that the expressions (\ref{ldfb}) and
(\ref{lamb}) are in accordance with the general properties of
$\mu(\lambda)$ discussed above. Finally, using (\ref{spc}) and
(\ref{pldf}) we obtain for the large deviation function $F(q)$,
\begin{eqnarray}
F(q)=-(1/2)\left[\Gamma_1+\Gamma_2-q(\lambda_++\lambda_-)-
(\lambda_+-\lambda_-) \sqrt{\Gamma_1\Gamma_2T_1T_2+q^2}\right],
\label{pldf2}
\end{eqnarray}
which yields a heat distribution $P(q)$ in accordance with the
general discussion in Sec.~\ref{discuss} A with exponential tails
in the heat distribution; for more details regarding this model,
see \cite{Fogedby11a}.
\subsection{Two particle chain}
\begin{figure}
\includegraphics[width=1.0\hsize]{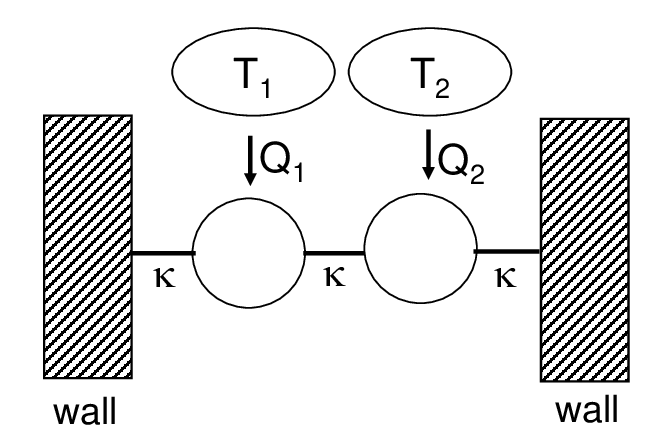}
\caption{We depict a chain composed of two particles interacting
with heat reservoirs at temperatures $T_1$ and $T_2$. The chain is
attached to walls or substrates at the ends. The heat transferred
to the particles is denoted $Q_1$ and $Q_2$, respectively. The
particle is attached to a substrate with a harmonic spring with
force constant $\kappa$. } \label{fig5}
\end{figure}

As an illustration of the general scheme presented here we briefly
consider the case of a chain composed of two particles; the
configuration is depicted in Fig.~\ref{fig5}. Setting $N=2$ we
have from (\ref{ff1}), (\ref{GN}), and (\ref{a2})
\begin{eqnarray}
&&|B(\omega)|^2=\frac{\kappa^2\omega^2}{|\Omega(\omega)^2-\kappa^2|^2},
\label{2a}
\\
&&\Omega(\omega)=-\omega^2+2\kappa-i\Gamma\omega, \label{2om}
\end{eqnarray}
and we obtain from the general expression (\ref{ldff}) the CGF
\begin{eqnarray}
\mu(\lambda)= -\frac{1}{2}\int\frac{d\omega}{2\pi} \ln\left[1+
\frac{4\Gamma^2\kappa^2\omega^2f(\lambda)}
{|\Omega(\omega)^2-\kappa^2|^2} \right]. \label{2ldfb}
\end{eqnarray}
We have been unable to reduce the expression (\ref{2ldfb}) further
but note that for $\kappa=0$ the LDF $\mu(\lambda)=0$ for all
$\lambda$, corresponding to two independent equilibrium systems at
temperatures $T_1$ and $T_2$. We also remark that decoupling the
chain from the walls, corresponding to setting $\Omega=-\omega^2
+\kappa-i\Gamma\omega$, we obtain
$4\Gamma^2\kappa^2\omega^2/(|\Omega^2-\kappa^2|^2) =
4\Gamma^2\kappa^2/((\omega^2+\Gamma^2)|\omega^2-2\kappa+i\Gamma\omega|^2)$.
In the limit of a stiff chain, corresponding to
$\kappa\rightarrow\infty$, we have
$4\Gamma^2\kappa^2\omega^2/(|\Omega^2-\kappa^2|^2)
\rightarrow\Gamma^2/(\omega^2+\Gamma^2)$, i.e., the case of a
single unbound particle coupled to two reservoirs, see
Sec.~\ref{discuss} B.
\subsection{$N$ particle chain}
In the limit of large $N$ we present below an asymptotic
expression for the CGF. For general $N$ we have from (\ref{dis}),
(\ref{GN}), (\ref{d}), and (\ref{a22})
\begin{eqnarray}
|B(\omega)|^2=\frac{4\kappa^3\sin^2(p/2)\sin^2(p)}
{|\Omega^2\sin(N-1)p-2\kappa\Omega\sin(N-2)p+\kappa^2\sin(N-3)p|^2},
\label{BN}
\end{eqnarray}
which expanding the denominator can be expressed in the form
\begin{eqnarray}
|B(\omega)|^2=\frac{8\kappa^{-1}\sin^2(p/2)\sin^2(p)}
{L(p)+K(p)\cos(2Np-\phi(p))}, \label{BN2}
\end{eqnarray}
where
\begin{eqnarray}
&&a=(1-\alpha^2)\cos(p)-2i\alpha,\label{a}
\\
&&b=(1+\alpha^2)\sin(p),\label{b}
\\
&&L=|a|^2+|b|^2, \label{L}
\\
&&M=|b|^2-|a|^2, \label{M}
\\
&&C=ab^\ast+a^\ast b, \label{C}
\\
&&K=\sqrt{M^2+C^2}, \label{K}
\\
&&\alpha=(2\Gamma/\sqrt\kappa)\sin(p/2),\label{al}
\\
&&\tan\phi=C/M.\label{tan}
\end{eqnarray}
By inspection we note that $|B(\omega)|^2$ displays an oscillatory
structure with approximate period $\pi/N$, reflecting the
resonance structure of the propagating lattice waves in the chain.
The oscillations are modulated by the slowly varying functions of
$p$, $\sin(p)$ and $\sin(p/2)$. Further inspection of (\ref{BN2})
shows that the maxima are given by
\begin{eqnarray}
|B(\omega)|_{\text{max}}^2=\frac{8\kappa^{-1}\sin^2(p/2)\sin^2(p)}
{L(p)-K(p)}, \label{BNmax}
\end{eqnarray}
where a little analysis implies that $|B(\omega)|_{\text{max}}^2$
locks onto $1/4\Gamma^2$, corroborating the demonstration of the
upper bound in appendix \ref{app1}. The lower bound of the
oscillatory structure is, correspondingly, given by the envelope
\begin{eqnarray}
|B(\omega)|_{\text{env}}^2=\frac{8\kappa^{-1}\sin^2(p/2)\sin^2(p)}
{L(p)+K(p)}, \label{BNenv}
\end{eqnarray}
the structure is for $N=10$ depicted in Fig.~\ref{fig9}. The
positions of the maxima and minima are given by the implicit
conditions $2Np-\phi(p)=\pi~~ (\text{mod}~ 2\pi)$ and
$2Np-\phi(p)=0~~ (\text{mod}~ 2\pi)$, respectively.

\begin{figure}
\center \psfrag{B}[ct][ct][1.5]{$|B(p)|^{2}$}
\psfrag{p}[ct][ct][1.5]{$p$}
\includegraphics[width=1.0\hsize]{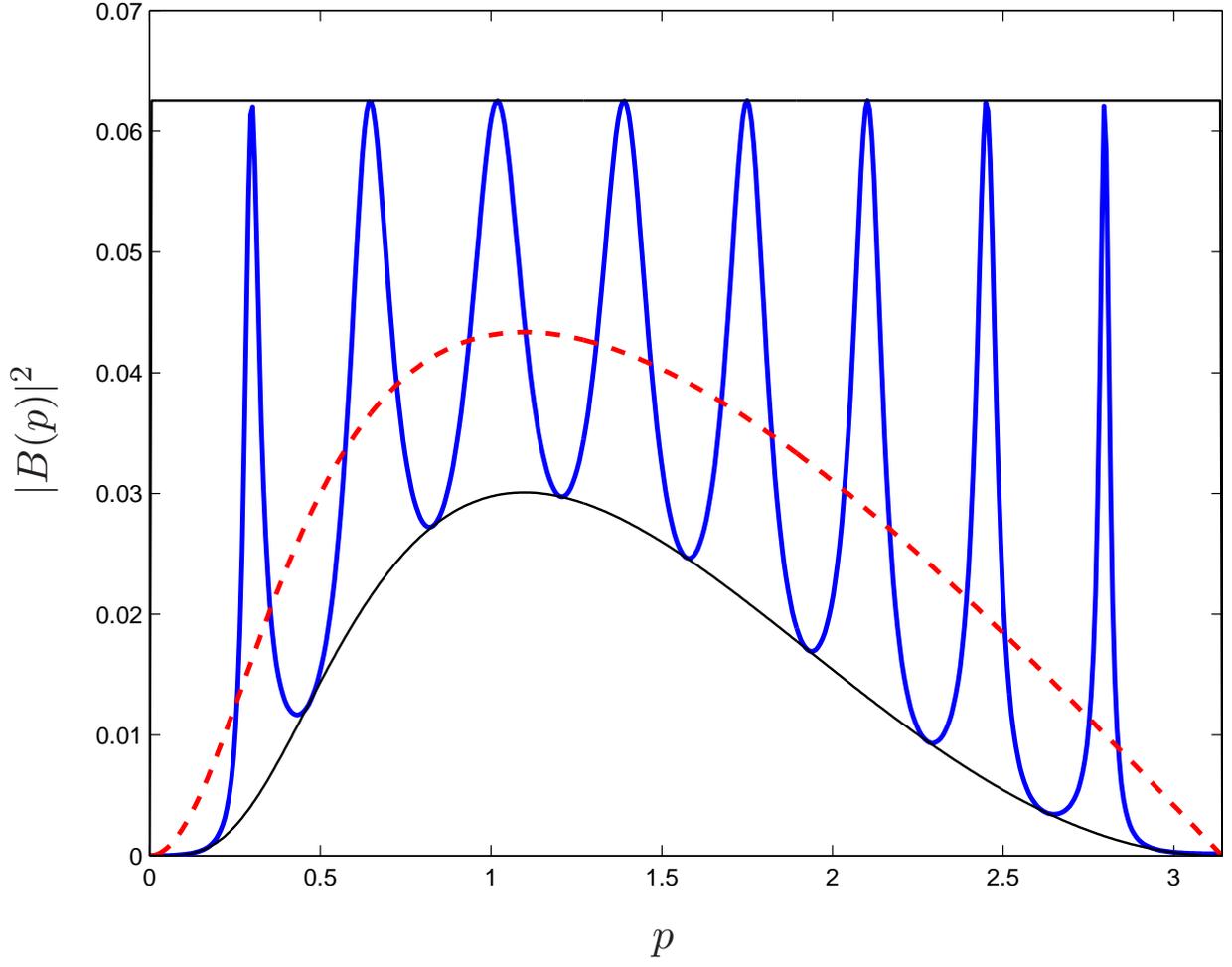}
\caption{We depict the squared modulus $|B|^2$ given by
(\ref{BN2}) as function of $p$ in the range $0<p<\pi$ for $N=10$,
$\Gamma=2$, and $\kappa=1$ (blue). We also show the maximum value
$|B|^2_{\text{max}}=1/4\Gamma^2$ (black) given by (\ref{BNmax}) ,
the envelope $|B|^2_{\text{env}}$ (black) given by (\ref{BNenv}),
and the large $N$ approximation $|B|^2_{\text{approx}}$ (red,
dashed) given by (\ref{BNapp}).} \label{fig6}
\end{figure}

\begin{figure}
\center\psfrag{m}[ct][ct][1.5]{$\bf\mu(\lambda)$}
\psfrag{l}[ct][ct][1.5]{$\lambda$}
\includegraphics[width=1.0\hsize]{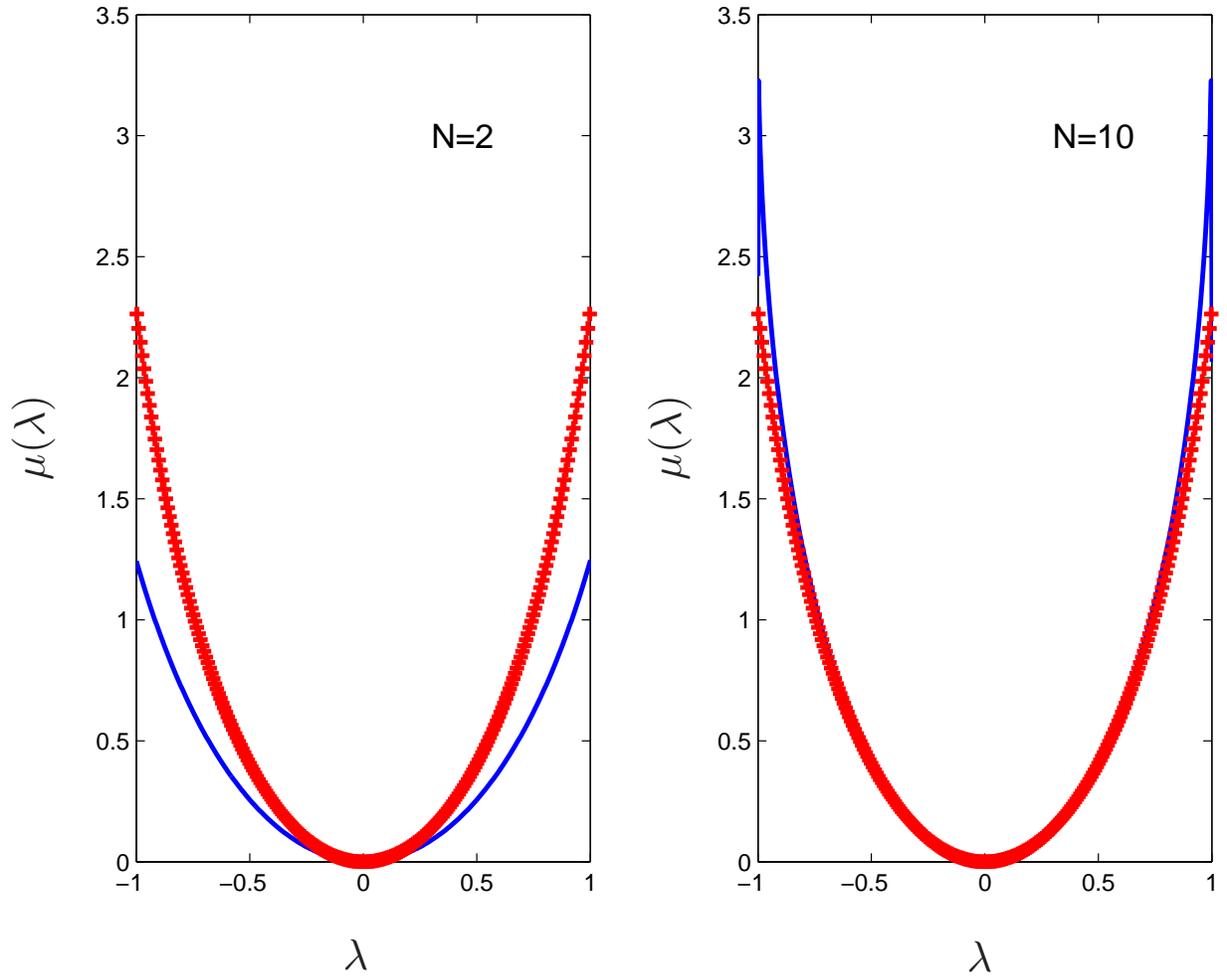}
\caption{We depict in two plots the LDF $\mu(\lambda)$ as a
function of $\mu$ for $N=2$ and $N=10$, respectively. The
parameters are $\Gamma=2$, $\kappa=1$, and $T_1=T_N=1$. The blue
curve is based on the exact expression given by (\ref{2ldfb2}),
the red plusses are given by the $N=\infty$ expression in
(\ref{2ldfb3}).} \label{fig7}
\end{figure}

In $p$ space, using $d\omega=\sqrt{\kappa}\cos(p/2)dp$, and
inserting (\ref{BN2}), the CGF given by (\ref{ldff}) takes the
form
\begin{eqnarray}
\mu(\lambda)= -\int_0^\pi\frac{dp}{2\pi}\sqrt{\kappa}\cos(p/2)
\ln\left[1+
\frac{16\Gamma^2\kappa^{-1}\sin^2(p/2)\sin^2(p)f(\lambda)}
{L(p)+K(p)\cos(2Np-\phi(p))} \right]. \label{2ldfb2}
\end{eqnarray}
In the large $N$ limit the rapid oscillations in $|B|^2$ allows us
to integrate separately over each period. Using the integral
\cite{Gradshteyn65}, see also ref. \cite{Roy08},
\begin{eqnarray}
\int_0^{2\pi}\frac{dp}{2\pi}\frac{1}{a+b\cos(p)}=\frac{1}
{\sqrt{a^2-b^2}}, \label{inte}
\end{eqnarray}
we thus obtain the following approximate form of $|B|^2$,
\begin{eqnarray}
|B|^2_{\text{approx}}= \frac{8\kappa^{-1}\sin^2(p/2)\sin^2(p)}
{\sqrt{L(p)^2-K(p)^2}}. \label{BNapp}
\end{eqnarray}
Further, inserting $L$ and $K$ from (\ref{L}) and (\ref{K}) we
obtain
\begin{eqnarray}
|B|^2_{\text{approx}}=\frac{2}
{\Gamma\sqrt\kappa}\frac{\sin(p/2)\sin(p)}
{1+4(\Gamma^2/\kappa)\sin^2(p/2)}, \label{BNapp2}
\end{eqnarray}
and for the CGF in the limit $N\rightarrow\infty$
\begin{eqnarray}
\mu(\lambda)= -\int_0^\pi\frac{dp}{2\pi}\sqrt{\kappa}\cos(p/2)
\ln\left[1+\frac{8\Gamma\kappa^{-1/2}\sin(p/2)\sin(p)f(\lambda)}{1+
4(\Gamma^2/\kappa)\sin^2(p/2)}\right]. \label{2ldfb3}
\end{eqnarray}
In Fig.~\ref{fig6} we have for $N=10$, $\Gamma=2$, and $\kappa=1$
depicted $|B|^2$, $|B|^2_{\text{max}}=1/4\Gamma^2$,
$|B|^2_{\text{env}}$, and $|B|^2_{\text{approx}}$. We note that
$|B|^2_{\text{approx}}$ smoothly interpolates over the
oscillations in $|B|^2$. In Fig.~\ref{fig7} we depict
$\mu(\lambda)$ as a function of $\mu$ for $N=2$ and for $N=10$.
The other parameters are $\Gamma=2$, $\kappa=1$, $T_1=1$, and
$T_N=1$. We note the excellent fit already for $N=10$ and the good
approximation at small $\lambda$ for $N=2$.

The expression  (\ref{2ldfb3}) for $\mu(\lambda)$ is manifestly
independent of $N$ in the large $N$ limit. This implies according
to (\ref{cum2}) that the cumulants and in particular the mean
current also are independent of $N$. This signals that Fourier's
law is not valid for the harmonic chain, see e.g. ref.
\cite{Rieder67}. We also note that the large $N$ limit does not
correspond to the continuum limit; we just increase the number of
particles in the chain keeping the lattice distance fixed.
\section{\label{gen} Generalized Fluctuation Theorem}
In Secs.~\ref{large} and ~\ref{discuss} we demonstrated the
validity of the fluctuation theorem by an explicit evaluation of
the CGF for the harmonic chain driven at the end points by heat
reservoirs at distinct temperatures and considered, moreover, some
special cases. Here we put these results in a more general
framework by considering the Fokker-Planck equation for the
characteristic function
\begin{eqnarray}
C(\lambda,t)=\langle e^{\lambda Q(t)}\rangle. \label{char2}
\end{eqnarray}
For long times $C(\lambda,t)\sim\exp(t\mu(\lambda))$ and we obtain
the differential equation
\begin{eqnarray}
\frac{\partial C}{\partial t}=\mu(\lambda)C. \label{diff}
\end{eqnarray}
Expressing the Fokker-Planck equation for $C$ in the form
\begin{eqnarray}
\frac{\partial C}{\partial t}=L(\lambda)C, \label{fp}
\end{eqnarray}
we identify the CGF $\mu(\lambda)$ as the maximal eigenvalue of
the Fokker-Planck operator $L$. The issue is thus to establish the
fluctuation theorem symmetry for the maximal eigenvalue.

\begin{figure}
\includegraphics[width=1.0\hsize]{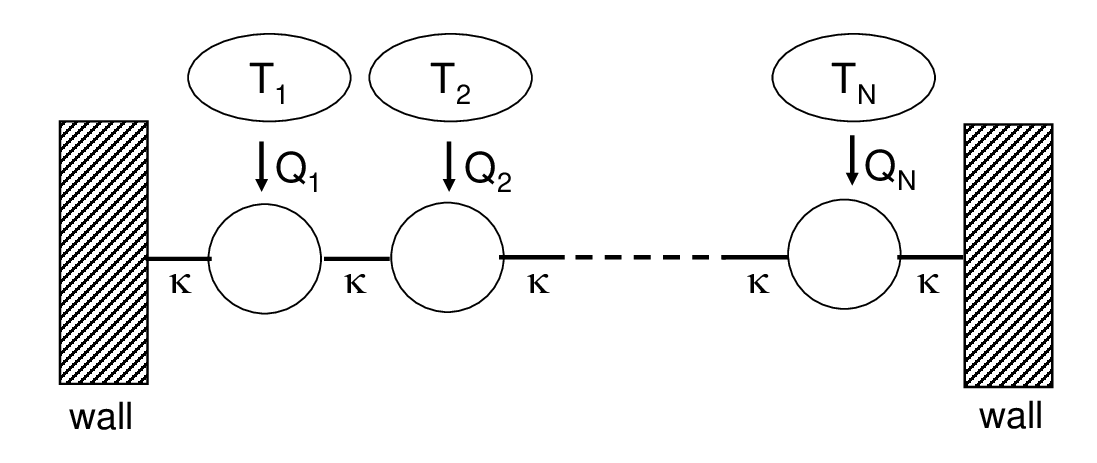}
\caption{We depict a harmonic chain where the n-th  particle is in
contact with a heat reservoir at temperatures $T_n$. The chain is
attached to walls or substrates at the ends. The total heat
transmitted to the n-th particle ise denoted $Q_n$ The spring
constant is denoted $\kappa$.} \label{fig8}
\end{figure}

We aim at a generalization of the fluctuation theorem to the case
of many heat reservoirs, see also \cite{Derrida05}. For that
purpose we consider a setup where each particle in the chain
couples to its own heat reservoir at temperature $T_n$. The
configuration is depicted in Fig.~\ref{fig8}. Generalizing
(\ref{hf}) the heat flux to the n-th particle is given by
\begin{eqnarray}
\frac{dQ_n}{dt}=p_n(-\Gamma p_n+\xi_n), \label{hfn}
\end{eqnarray}
where the noise is correlated according to
\begin{eqnarray}
\langle\xi_n(t)\xi_m(t')\rangle=2\delta_{nm}\Gamma
T_n\delta(t-t'). \label{nnm}
\end{eqnarray}
Since the transfer of heat induces a change in the state of the
system we must at the outset consider the joint distribution
$P(u,p,Q,t)\equiv P(\{u_n\},\{p_n\},\{Q_n\},t)$. The heat
distribution is then given by $P(Q,t)=
\int\prod_ndu_ndp_nP(u,p,Q,t)$.

As discussed in ref. \cite{Imparato07}, the Fokker-Planck equation
for the joint distribution $P(u,p,Q,t)$ is derived by considering
the heat $Q_n(t)$ as an independent dynamical variable, whose time
evolution is governed by (\ref{hfn}). Noting that the noise
appearing in this equation is correlated to the noise appearing in
the equation of motion for the momenta (\ref{eq1}-\ref{eq4}) one
can write
\begin{eqnarray}
\frac{\partial P}{\partial t}= &&\{P,H\}+
\Gamma\sum_n\left(T_n\frac{\partial^2P}{\partial
p_n^2}+\frac{\partial}{\partial p_n}(p_nP)\right) \nonumber
\\
+&&\Gamma\sum_n\left( \frac{\partial}{\partial Q_n}((p_n^2+T_n)P)+
T_np_n^2\frac{\partial ^2P}{\partial Q_n}+2T_n
p_n\frac{\partial^2P}{\partial Q_n\partial p_n}\right),
\label{fp2}
\end{eqnarray}
where the Poisson bracket is given by
\begin{eqnarray}
\{P,H\}=\sum_{n=1}^N\left[\frac{\partial P}{\partial
p_n}\frac{\partial H}{\partial u_n}-\frac{\partial P}{\partial
u_n}\frac{\partial H}{\partial p_n} \right]; \label{pb}
\end{eqnarray}
see also ref. \cite{Lebowitz99}.

All reference to the deterministic dynamics of the chain is
embodied in the Poisson bracket. The remaining terms in
(\ref{fp2}) are associated with the transfer of heat. Setting
$\partial/\partial Q_n=-\lambda_n$ and $\partial ^2/\partial
Q_n^2=\lambda_n^2$ we obtain for the characteristic function
$C(u,p,\{\lambda_n\},t)$ defined by the multiple Laplace transform
\cite{Gradshteyn65}
\begin{eqnarray}
P(u,p,\{Q_n\},t)=\int_{-i\infty}^{i\infty}\prod_n\frac{d\lambda_n}{2\pi
i} \exp(-\sum_n \lambda_n Q_n)C(u,p,\{\lambda_n\},t),
\end{eqnarray}
the Fokker-Planck equation (\ref{fp}), where the operator
$L(\lambda)$ has the form
\begin{eqnarray}
L(\lambda)C=&&\{C,H\}
\nonumber
\\
+&&\Gamma\sum_n\left[T_n\frac{\partial^2C}{\partial p_n^2}+
(1-2\lambda_n T_n)\frac{\partial}{\partial p_n}(p_nC) +
(\lambda_n(\lambda_nT_n-1)p_n^2+\lambda_nT_n)C\right].~~~~
\label{fp4}
\end{eqnarray}
In the absence of coupling between the particles, i.e., for a
vanishing Poisson bracket, $\{C,H\}=0$, $C$ is the characteristic
function for the the heat transfers to $N$ independent particles
coupled individually to reservoirs at temperature $T_n$. Subject
to the transformation $C=\exp(Et)\exp(g)\Psi$, where
$g=(1/2)\sum_n(\lambda_n-1/2T_n)p_n^2$, the Schroedinger-like
equation $L\Psi=E\Psi$ describes $N$ independent oscillators with
spectrum $E=-\Gamma(n_1+n_2+\cdots n_N), n_i=0,1,\cdots$. The
maximal eigenvalue is given by $E=0$ corresponding to $\mu=0$,
characteristic of an equilibrium configuration. Turning on the
interaction between the particles the maximal eigenvalue will be
shifted to a finite value and we obtain a nonvanishing
$\lambda$-dependent CGF.

The structure of (\ref{fp4}) also allows a simple derivation of a
generalized fluctuation theorem; see also ref. \cite{Derrida05}.
The first step is to perform a ``rotation" $\exp(H/T_m)$ with
respect to the m-th reservoir in combination with a time reversal
operator $\cal T$ and define the transformed Fokker-Planck
operator
\begin{eqnarray}
\tilde L(\lambda)= e^{H/T_m}{\cal T}L(\lambda){\cal
T}^{-1}e^{-H/T_m}. \label{fpop2}
\end{eqnarray}
In the next step we compare the operator $\tilde L$ with the
adjoint operator $L^\ast$. Using $(\partial^2/\partial
p_n^2)^\ast=\partial^2/\partial p_n^2$ and $(\partial p_n/\partial
p_n)^\ast=-p_n\partial /\partial p_n$ and shifting the Laplace
variables $\lambda_n$ it turns out that $\tilde L$ and $L^\ast$
become identical and we have the relationship
\begin{eqnarray}
\tilde L(\lambda)=L^\ast(\bar\lambda), \label{op}
\end{eqnarray}
where
\begin{eqnarray}
\bar\lambda_n+\lambda_n=1/T_n-1/T_m. \label{lam}
\end{eqnarray}
Since $L(\lambda)$ is related to  $\tilde L(\lambda)$ by a unitary
transformation we infer that $L(\lambda)$ and
$L^\ast(\bar\lambda)$ have identical spectra and in particular
identical maximal eigenvalues, i.e., the same large deviation
function,
\begin{eqnarray}
\mu(\{\lambda_n\})=\mu(\{\bar\lambda_n\}).  \label{ftg}
\end{eqnarray}
The expression (\ref{ftg}) together with (\ref{lam}) represents a
generalization of the usual fluctuation theorem to many
reservoirs. In the case of two reservoirs, setting $T_n=T_1$,
$T_m=T_N$, and $\lambda_n=\lambda$ we obtain the usual fluctuation
theorem (\ref{ft}). We note that the above derivation holds for
any time reversal invariant Hamiltonian, i.e., for any kind of
interaction between the particles. In our derivation we have also
assumed that the maximal eigenvalue is positive.

Thus, our proof of the fluctuation theorem is more general than
the one given in Ref.~\cite{Saito10b}, which is restricted to the
harmonic chain only. Furthermore, our approach does not require a
direct evaluation of the CGF, but is based only on the property of
the dynamics, as expressed by the evolution operator $L(\lambda)$.
The previous proof can be readily extended to the 3-D case, as
long as  $L(\lambda)$ has a form as in (\ref{fp4}).


\section{\label{sum} Summary and conclusion}
In this paper we have discussed a variety of issues regarding the
noise driven harmonic chain. In Secs.~\ref{analysis} and
~\ref{large} we performed a calculation of the CGF, recovering the
results of Kundu et al. \cite{Kundu11}, but adding some more
details for the purpose of our analysis. In Sec~\ref{discuss} we
discussed the exponential tails in the heat distribution, the
bound single particle model, and the two-particle chain case. It
is an interesting feature of the tails that the fall-off rate only
depends on the noise features, i.e., the reservoir temperatures,
and not on the dynamical properties of the chain such as the
spring constant $\kappa$. In the large $N$ limit we have found an
analytical form for the CGF which excellently interpolates the
exact result. This result is independent of $N$ signalling that
Fourier'r law does not hold. Finally, incorporating some of our
results we have in Sec.~\ref{gen} within a Fokker-Planck
description presented a generalization of fluctuation theorem to
several reservoirs which holds for any interaction potential. The
fluctuation theorem simply emerges from a symmetry hidden in the
Fokker- Planck operator and is therefore not restricted to the
linear chain but also holds for a 3D system.

\acknowledgements We are grateful to C. Mejia-Monasterio and F.
van Wijland for interesting discussions. We also thank A. Mossa,
A. Svane, U. Poulsen, and G. Bruun for useful discussions. HF
gratefully acknowledges financial support by the Danish Natural
Science Research Council under grant no. 09-072352. AI gratefully
acknowledges financial support by Lundbeck Fonden.

\appendix
\section{Maxima of $|B(\omega)|^2$}
\label{app1}

Here we analyze the modulus squared of the function
\begin{equation}
B(\omega)=\frac{-i \omega \kappa \sin p}{D(\omega)},
\label{a2:app}
\end{equation}
where
\begin{eqnarray}
&&D(\omega)=
\Omega^2\sin(N-1)p-2\kappa\Omega\sin(N-2)p+\kappa^2\sin(N-3)p,\\
&&\Omega = -\omega^2+2\kappa-i\Gamma\omega,\\
&&\omega^2= 4 k \sin^2 (p/2), \label{omegap}
\end{eqnarray}
and demonstrate that it is bounded from above by $1/4\Gamma^2$,
i.e.,$|B(\omega)|^2 \le 1/(4 \Gamma^2)$.

\begin{figure}
\center \psfrag{A24G2}[ct][ct][1.]{$|B(p)|^{2}  (4 \Gamma^2)$}
\psfrag{p}[ct][ct][1.]{$p$}
\includegraphics[width=15cm]{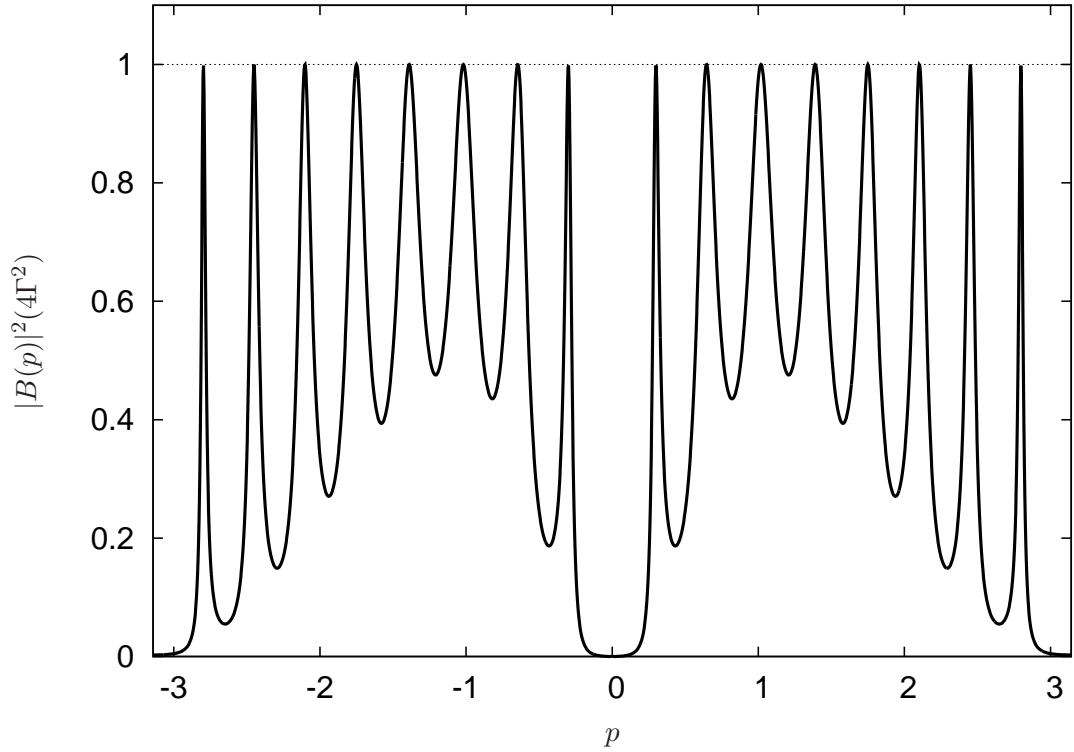}
\caption{Plot of the squared modulus of the momentum Green's
function $B$ as a function of $p$, as given by (\ref{a2:app}) for
$\Gamma=2$, $\kappa=1$, $N=10$.}\label{fig9}
\end{figure}

Breaking up $|D(\omega)|$ in real and imaginary parts,
\begin{eqnarray}
\Re\pq{D(\omega)}&=&-4 \kappa \Gamma^2 \sin^2(p/2) \sin(N-1)p+
\kappa^2 \sin(N+1)p, \label{red}
\\
\Im\pq{D(\omega)}&=&-4 \kappa^{3/2}\Gamma \sin(p/2) \sin N p,
\label{imd}
\end{eqnarray}
inserting $|B(\omega)|^2$, expressing $\omega$ in terms of $p$,
using (\ref{omegap}), and substituting (\ref{red})-(\ref{imd}), we
rephrase the condition $|B(p)|^2 \le 1/(4 \Gamma^2)$ as
\begin{eqnarray}
g(p)&=&16\kappa^3 \Gamma^2 \sin^2(p/2) \pq{\sin^2 N p-\sin^2 p}
\nonumber
\\
&&+(-4\kappa\Gamma^2\sin(p/2)\sin(N-1)p+\kappa^2\sin(N+1)p)^2 \ge
0.
\end{eqnarray}
Expressing the $\sin(p/2)$ in terms of $\cos p$, and rearranging
terms, $g(p)$ becomes
\begin{equation}
g(p)=k^2 (-\Gamma^2 \sin Np+2 \Gamma^2 \sin(N-1)p-\Gamma^2
\sin(N-2)p+ k\sin(N+1) p)^2,
\end{equation}
which is non negative thus demonstrating our assertion. The values
of $p$ for which $g(p)=0$ correspond to the points of maximum for
$|B(\omega)|^2$, with the exception of $p=0,\, \pi$ where
$D(\omega)=0$. So, $|B(\omega)|^2$ has N-1 maxima where
$|B(\omega)|^2=1/(4 \Gamma^2)$, see Fig.~\ref{fig9}.


\newpage

\end{document}